\documentclass[published]{JHEP}

\usepackage{amssymb}

\newcommand{\md}{{\rm d}}
\newcommand{\e}{{\rm e}}
\newcommand{\T}{{\rm T}}
\newcommand{\eq}[1]{eq.~(\ref{#1})}
\newcommand{\bra}[1]{\langle #1 |}
\newcommand{\ket}[1]{| #1 \rangle}
\newcommand{\mrm}[1]{\mathrm{#1}}
\newcommand{\msf}[1]{\mathsf{#1}}

\title{Multi-channel Bethe-Salpeter equation\thanks{This work
has been supported in part under the German-Polish agreement on
bilateral scientific and technological cooperation.}}

\author{Jaroslaw Boguszy\'nski\\ State Office for Insurance Supervision,
Warsaw, Poland\\ E-mail: \email{jaroslaw.boguszynski@fuw.edu.pl}}

\author{Hans Dieter Dahmen and Ralph Kretschmer\\ Siegen University, Fachbereich
Physik, Siegen, Germany\\ E-mail:
\email{dahmen@physik.uni-siegen.de},
\email{kretschm@hepth2.physik.uni-siegen.de}}

\author{Leszek {\L}ukaszuk\\ Soltan Institute for Nuclear Studies,
Warsaw, Poland\\ E-mail: \email{leszek.lukaszuk@fuw.edu.pl}}

\abstract{A general form of multi-channel Bethe-Salpeter equation is
considered. In contradistinction to the hitherto applied approaches,
our coupled system of equations leads to the simultaneous solutions
for all relativistic four-point Green functions (elastic and
inelastic) appearing in a given theory. A set of relations which may
be helpful in approximate treatments is given. An example of
extracting useful information from the equations is discussed: we
consider the most general trilinear coupling of $N$ different scalar
fields and obtain --- in the ladder approximation --- closed
expressions for the Regge trajectories and their couplings to
different channels in the vicinity of $\ell = -1$. Sum rules and an
example containing non-obvious symmetry are~discussed.}

\keywords{Phenomenological Models, Sum Rules, Integrable Equations in Physics, Nonperturbative Effects}

\received{May 16, 2000}

\accepted{November 01, 2000\smash{\llap{\raise-1\baselineskip
\hbox{{\sc Revised:} October 05, 2000}}}}

\JHEP{11(2000)002}

\begin{document}

\section{Introduction}\label{s1}

The Bethe-Salpeter (BS) equation~\cite{Bethe} has been widely applied
in studies of two-body scattering and bound-state problems (reviews
can be found in~\cite{Nakanishi1}). It is still the most important
tool for describing bound states in quantum field theories. Recent
applications include studies of quark-antiquark bound states by a
combination of the BS equation and the Dyson-Schwinger equations in
four-dimensional euclidean space~\cite{Jain...,Roberts}. In addition,
three-dimensional reductions of the BS equation are the basis of
models with effective potentials that also lead to surprisingly good
predictions~\cite{Lucha...}. Studies of the BS equation in
four-dimensional Minkowski space on the basis of the perturbation
theory integral representation~\cite{Nakanishi2}, which avoid the
difficulties of performing a Wick rotation~\cite{Kusaka}, or attempts
go beyond the ladder approximation~\cite{Nieuwenhuis} have lead to
interesting results.

In some of the recent applications of the BS equation to
scattering processes, the importance of a consistent
treatment of coupled channels has been emphasized. Oller
et al.\ \cite{Oller} and Oset et al.\ \cite{Oset...}
have developed a method that combines a multi-channel BS
equation with chiral perturbation theory
(ChPT)~\cite{Pich} to describe scalar
mesons~\cite{Oller} and meson-nucleon
scattering~\cite{Oset...}. Here, the interest in the
interpretation of scalar resonances stems from the
possibility \pagebreak[3] that some of these resonances
may be glueballs or at least mixtures of $q \bar{q}$ and
gluonic states~\cite{Close}. In this context one should
also mention the works of Janssen et al.\ \cite{Janssen}
and Kami\'{n}ski et al.\ \cite{Kaminski}, who employ
coupled-channel versions of the Lippmann-Schwinger
equation. For the description of meson-nucleon
scattering, the inclusion of additional channels widens
the range of applicability of the lowest order chiral
lagrangian~\cite{Kaiser,Oset...}.

Although these applications of the BS equation are
phenomenologically quite successful, none of them provides a
fully relativistic description. In a one-channel context, fully
relativistic calculations were carried out for meson-meson
scattering in~\cite{Nieves} and for meson-nucleon scattering
in~\cite{Lahiff}.

But the BS equation is also a valuable tool in axiomatic
QFT: it has been demonstrated (\cite{Bros1,Bros2} and
references therein) that the exact Bethe-Salpeter
formalism in the complex momentum space --- together
with its analytical dependence on both complex energy
and complex angular momentum --- follows from the basic
principles of massive QFT such as locality, Lorentz
covariance and the spectral condition. The possibility
of an extension of this rigorous approach to theories of
QCD type with a discrete spectrum of composite particles
appearing as Regge-type particles has been recently
announced~\cite{Bros2}.

Having in mind these impressive efforts and developments it seems
rather surprising that the general form of the BS equation for
two-body multi-channel problems has not been discussed in the
literature. We find it therefore appropriate to start such a
discussion.

Our contribution to this subject will be an
investigation of some general properties of the
solutions, leading to decomposition formulas for the
case when the Bethe-Salpeter kernel and/or the Green
functions can be subdivided into simpler parts. As an
illustration of the usefulness of the multi-channel
formulation, we then calculate the positions and
couplings of Regge poles in the vicinity of $\ell = - 1$
in a theory with general trilinear coupling of $N$
different scalar fields in the ladder approximation. The
treatment of this example is motivated not only by the
fact that scalar theories of this kind, e.g.\ the
Wick-Cutkosky model or the fully massive $\varphi^2
\sigma$ model are widely used~\cite{Kusaka,Nieuwenhuis},
but also because studies of the $\ell = - 1$ singularity
in $\varphi^3$ theory often serve as a starting point
for investigations of the diffractive region in QCD (see
e.g.~\cite{Bartels}).

As a result we obtain some novel features of Regge behaviour, in
particular we derive non-trivial sum rules for the scattering
amplitudes in the different channels.

The multi-channel approach presented here may be helpful
to solve the important problem of understanding
multi-channel hadron-hadron scattering as a consequence
of QCD. There exist many approaches which could be a
good point of departure for multi-channel BS
predictions. What we have in mind are successful
theories like ChPT~\cite{Pich}, where effective actions
with composite, hadronic degrees of freedom are derived,
e.g.~\cite{Bando,Cahill...}. Among these, the Global
Colour Model described in~\cite{Cahill...,Roberts}
respects the ultraviolet behaviour of QCD so that
natural cutoffs appear in the effective hadronic
couplings. In this context our example of a
super-renormalizable theory with many channels involved
may exhibit essential features of more realistic, future
applications.

\section{The symmetric and reduced forms of the Bethe-Salpeter
equation}\label{s2}

Consider the non-truncated Bethe-Salpeter Green function
$\tau_\mrm{BS}$:
\begin{eqnarray}
\tau_\mrm{BS}(a, x_1; b, x_2| a', x_1'; b'; x_2')
&=& \bra{0} \T [ \varphi_a(x_1) \varphi_b(x_2) \varphi_{a'}(x_1')
\varphi_{b'}(x_2')] \ket{0}- \label{2l} \\
&&-\, \bra{0} \T [\varphi_a(x_1) \varphi_b(x_2)] \ket{0} \bra{0}
\T [\varphi_{a'}(x_1') \varphi_{b'}(x_2')] \ket{0}\,. \nonumber
\end{eqnarray}
Here the fields $\varphi_a(x)$ may be bosons, fermions or ghosts
--- formally, we use one multicomponent field and the indices $a$
contain the information on its boson, fermion or ghost nature. The
function $\tau_\mrm{BS}$ has the following symmetries:
\begin{eqnarray}
\tau_\mrm{BS}(a, x_1; b, x_2| a', x_1'; b', x_2')
&=& \delta^\mrm{P}_{a b} \,
\tau_\mrm{BS}(b, x_2; a, x_1| a', x_1'; b', x_2') \label{2d} \\
&=& \delta^\mrm{P}_{a' b'} \,
\tau_\mrm{BS}(a, x_1; b, x_2| b', x_2'; a', x_1')\,, \\
\tau_\mrm{BS}(a, x_1; b, x_2| a', x_1'; b', x_2')
&=& \delta^\mrm{P}_{a a'} \, \delta^\mrm{P}_{a b'} \,
\delta^\mrm{P}_{b a'} \, \delta^\mrm{P}_{b b'} \,
\tau_\mrm{BS}(a', x_1'; b', x_2'| a, x_1; b, x_2)\,,
\label{2f}
\end{eqnarray}
where $\delta^\mrm{P}_{ab} = - 1$ for $\varphi_a$, $\varphi_b$
both being fermion (or ghost) fields, otherwise it is $+ 1$. The
symmetry~(\ref{2f}) results from the fact that the full propagator
\begin{equation}\label{2m}
\Delta'(a, x_1; b, x_2) = \delta^\mrm{P}_{ab} \, \Delta'(b, x_2;
a, x_1) = \bra{0} \T [\varphi_a(x_1) \varphi_b(x_2)] \ket{0}
\end{equation}
in the subtracted term in~(\ref{2l}) vanishes unless both
$\varphi_a$ and $\varphi_b$ are either both fermions or bosons.
But besides this, the propagator does not need to be diagonal in
the discrete indices. In such a manner we admit both spinor
propagators and the eventual mixing of different elementary
fields. With~(\ref{2m}) we can relate the matrix for the
truncated, connected 4-point function
\begin{equation}
M(a, x_1; b, x_2|a', x_1'; b', x_2') = \left. \bra{0} \T
[\varphi_a(x_1) \varphi_b(x_2) \varphi_{a'}(x_1')
\varphi_{b'}(x_2')] \ket{0} \right|_\mrm{trunc,\, conn}
\end{equation}
to $\tau_\mrm{BS}$ through
\begin{equation}\label{2a}
\tau_\mrm{BS} = G + G \bullet M \bullet G\,,
\end{equation}
where
\begin{eqnarray}
G(a, x_1; b, x_2| a', x_1'; b', x_2')&=& \delta^\mrm{P}_{a b} \,
\Delta'(a, x_1; a', x_1') \Delta'(b, x_2; b', x_2')+\nonumber\\
&&+\, \Delta'(a, x_1; b', x_2') \Delta'(b, x_2; a_1', x_1')\,,
\end{eqnarray}
and the $\bullet$ product is defined as
\begin{eqnarray}
(G \bullet M)(a, x_1; b, x_2| a', x_1'; b', x_2')
&=& \frac{1}{2} \sum_{c, d} \int \md^4 y_1 \, \md^4 y_2 \,
G(a, x_1; b, x_2| c, y_1; d, y_2)\times\nonumber\\
&&\hphantom{\frac{1}{2} \sum_{c, d} \int}\!\times M(c, y_1; d,
y_2| a', x_1'; b', x_2')\,. \label{2c}
\end{eqnarray}
The factor $1 / 2$ in~(\ref{2c}) appears because $G$ and $M$ fulfill
the same symmetry relations~(\ref{2d})--(\ref{2f}) as
$\tau_\mrm{BS}$ and hence all terms in the sum are counted twice. If
the momentum-space Green functions are normalized like
\begin{eqnarray}
M(a, k_1; b, k_2| a', k_1'; b', k_2') &=& \int \md^4 x_1 \,
\md^4 x_2 \, \md^4 x_1' \, \md^4 x_2' \, \e^{i (k_1 x_1 + k_2 x_2
- k_1' x_1' - k_2' x_2')}\times\nonumber\\
&&\hphantom{\int}\!\times M(a, x_1; b, x_2| a', x_1'; b', x_2')\,,
\end{eqnarray}
then the matricial form~(\ref{2a}) can also be used in 4-momentum
space, with the symbol $\bullet$ denoting
\begin{equation}
\bullet = \frac{1}{2} \sum_{c, d} \int \frac{\md^4 q_1}{(2 \pi)^4}
\frac{\md^4 q_2}{(2 \pi)^4}\,.
\end{equation}
In momentum space, $M$ contains an overall $\delta$ function,
\begin{eqnarray}
i M(a, k_1; b, k_2| a', k_1'; b', k_2')
&=& (2 \pi)^4 \delta^4(k_1 + k_2 - k_1' - k_2')\times\nonumber\\
&&\times\,\mathcal{M}(a, k_1; b, k_2| a', k_1'; b', k_2')\,,
\end{eqnarray}
and if the transition matrix is defined by $S = 1 - i T$, then
$\mathcal{M}$, taken at $k_1' + k_2' = k_1 + k_2$, denotes an
off-shell extrapolation of the scattering amplitude for the
process with outgoing particles $a$ and $b$ with momenta $k_1$
and $k_2$, respectively, and incoming particles $a'(k_1')$ and
$b'(k_2')$. In the same way, $G$ in momentum space contains
$\delta$~functions:
\begin{eqnarray}
G(a, k_1; b, k_2| a', k_1'; b', k_2')
&=& (2 \pi)^8 \left( \delta^\mrm{P}_{a b} \Delta'(a; a', k_1) \Delta'(b; b', k_2)
\delta^4(k_1 - k_1') \delta^4(k_2 - k_2') +\right. \nonumber \\
&& \left. +\, \Delta'(a; b', k_1) \Delta'(b; a', k_2)
\delta^4(k_1 - k_2') \delta^4(k_2 - k_1') \right)\label{2i}
\end{eqnarray}
with
$\Delta'(a; b, k_1)
= \int \md^4 x \, \e^{i k_1 x} \Delta'(a, x; b, 0)$.

\subsection{The symmetric form of the BS equation}

The matrices $\tau_\mrm{BS}$, $G$ and $M$ fulfill the symmetry
relations~(\ref{2d})--(\ref{2f}). This leads to the symmetric form
of the Bethe-Salpeter equation:
\begin{equation}\label{2b}
\tau_\mrm{BS} = G + G \bullet B \bullet \tau_\mrm{BS}\,,
\end{equation}
and
\begin{equation}\label{2j}
M = B + B \bullet G \bullet M\,.
\end{equation}
In these equations, $B$ is the two-particle irreducible (for any
two particles involved) part of $M$ and also possesses the
symmetries~(\ref{2d})--(\ref{2f}). The symmetry~(\ref{2f})
leads to the alternative forms
\begin{equation}\label{2g}
\tau_\mrm{BS} = G + \tau_\mrm{BS} \bullet B \bullet G
\qquad \mbox{and} \qquad
M = B + M \bullet G \bullet B
\end{equation}
of the BS equation. If we symbolically denote the solution
of~(\ref{2j}) by $M = M\{B, G\}$, then the solution of~(\ref{2b})
can be written analogously:
\begin{equation}
M = M\{B, G\}\,,\qquad \tau_\mrm{BS} = M\{G, B\}\,.
\end{equation}
This will be useful for the discussion of the general relations.

\subsection{The reduced form of the BS equation}

In applications one may encounter forms of the BS
equation that differ from the symmetric ones~(\ref{2b})
and~(\ref{2j}). These are special cases of a reduced
form of the BS equation, i.e.\ forms where the BS
equation is simplified for the price of loosing the
explicitly symmetric shape.

The general transition may be described as follows: let
$G^\mrm{r}$ and $B^\mrm{r}$ be arbitrary but satisfying
\begin{eqnarray}
G &=& G^\mrm{r}(a, x_1; b, x_2| a', x_1'; b', x_2')
+ \delta^\mrm{P}_{a' b'} \,
G^\mrm{r}(a, x_1; b, x_2| b', x_2'; a', x_1')\,,\nonumber \\
B &=& B^\mrm{r}(a, x_1; b, x_2| a', x_1'; b', x_2')
+ \delta^\mrm{P}_{a' b'} \,
B^\mrm{r}(a, x_1; b, x_2| b', x_2'; a', x_1')\,.
\end{eqnarray}
Then~(\ref{2b}) and~(\ref{2j}) can be transformed into
\begin{equation}\label{2h}
\tau_\mrm{BS} = G + G^\mrm{r} \circ B^\mrm{r} \circ
\tau_\mrm{BS}\,,\qquad M = B + B^\mrm{r} \circ G^\mrm{r} \circ M
\end{equation}
and similarly for~(\ref{2g}), with the $\circ$ product now being
defined as
\begin{equation}
\circ = \sum_{c, d} \int \md^4 y_1 \, \md^4 y_2 \,,
\end{equation}
i.e.\ without the factor $1 / 2$ of~(\ref{2c}). The
forms~(\ref{2h}) do not fit into the general relations~(\ref{2b})
and~(\ref{2j}). However, looking into the iteration procedure, we
find
\begin{equation}
\tau_\mrm{BS} = \frac{1}{1 - G^\mrm{r} \circ B^\mrm{r}}
\circ G =
\tau_\mrm{BS}^\mrm{r} + \delta^\mrm{P}_{a' b'}
\tau_\mrm{BS}^\mrm{r}\left((a', x_1') \leftrightarrow
(b', x_2')\right)
\end{equation}
with
\begin{equation}
\tau_\mrm{BS}^\mrm{r} = \frac{1}{1 - G^\mrm{r} \circ B^\mrm{r}}
\circ G^\mrm{r}\,.
\end{equation}
Thus, we can deal with the equation
\begin{equation}\label{2k}
\tau_\mrm{BS}^\mrm{r} = G^\mrm{r} + G^\mrm{r} \circ B^\mrm{r}
\circ \tau_\mrm{BS}^\mrm{r}\,.
\end{equation}
Similarly, one can decompose $M = M^\mrm{r} + \delta^\mrm{P}_{a'
b'} M^\mrm{r}((a', x_1') \leftrightarrow (b', x_2'))$ to work with
\begin{equation}\label{2e}
M^\mrm{r} = \frac{1}{1 - B^\mrm{r} \circ G^\mrm{r}} \circ B^\mrm{r}
= B^\mrm{r} + B^\mrm{r} \circ G^\mrm{r} \circ M^\mrm{r}\,.
\end{equation}
The functions $G^\mrm{r}$ and $B^\mrm{r}$ are not uniquely
determined. Let us add that in typical ladder approximations the
reduced form of the BS equation is in fact used. In these cases
the choice of suitable $G^\mrm{r}$ and $B^\mrm{r}$ has been
always implicitly made.

\section{General relations}\label{s3}

The basic principles of QFT ensure~\cite{Bros1,Bros2} that the
scattering matrix function $M$ in the physical region can be
obtained from the solution of \eq{2j} in the euclidean region by
analytic continuation in the invariant variables. Thus, if we had
complete knowledge of the matrix functions $B$ and $G$ in the
euclidean region, and if we would be able to exactly solve
\eq{2j} in this region, we could obtain complete information on
$M$ in the physical region. Generally none of these conditions is
met and one has to rely on soundness of chosen scheme of
approximations. Having this in mind we shall write below a few
general relations which in our opinion may be helpful at least as
a consistency check of approximations. We will throughout this
section use the notation~(\ref{2b}) and~(\ref{2j}) for the
symmetric form of the BS equation, but in view of~(\ref{2k})
and~(\ref{2e}) these relations can of course also be used for the
reduced forms.

To begin with, consider the solution of~(\ref{2j}) for given $B$,
$G$:
\begin{equation}\label{3d}
M \{B, G \} = \frac{1}{1 - B \bullet G} \bullet B
= B \bullet \frac{1}{1 - G \bullet B}\,.
\end{equation}
This solution, treated as a functional of $B$ and $G$ has the
following interesting property: for arbitrary $G_{1}$ one has
\begin{equation}\label{3f}
M \{B, G \} = M \left\{M \{B, G-G_{1} \}, G_{1}
\right\}.
\end{equation}
This equality has been found useful for the one-channel problem
long time ago~\cite{Yaes...}. In fact eq.~(\ref{3f}) comes
as an answer to the following question: what is the relation
between pairs $(B, G)$ and $(B_{1}, G_{1})$ if both of them lead
to the same $M$:
\begin{equation}
M \{B, G \} = M \{B_{1}, G_{1} \}\,.\label{3g}
\end{equation}
Using eqs.~(\ref{3d}) and~(\ref{3g}) we get $B_{1} = B + B
\bullet [G - G_{1}] \bullet B_{1}$, i.e.\ from~(\ref{2j}) we have
$B_{1} = M \{ B, G - G_{1} \}$; inserting it into~(\ref{3g}) we
end up with relation~(\ref{3f}).

Equation~(\ref{3f}) leads to the following formula for $G = G_{1}
+ G_{2} + G_{3} + \cdots + G_{n}$:
\begin{equation}
M \{B, G \} = M \left\{\dots M\{M\{M\{B,G_{1}\},
G_{2}\}, G_{3}
\},\dots, G_{n}\right\}\,. \label{3e}
\end{equation}

It is natural to ask whether similar formula holds in a situation
when we subdivide $B$ instead of $G$. The answer is yes, once we
notice relation~(\ref{2a}) between the off-shell scattering
amplitude $M\{B, G \}$ and the non-truncated 4-point Green
function $\tau_\mrm{BS} = M\{G, B \}$:
\begin{equation}\label{3a}
M \{B, G \} = B + B \bullet M \{G, B \} \bullet B\,,\qquad
M \{G, B \} = G + G \bullet M \{B, G \} \bullet G\,.
\end{equation}
Inserting~(\ref{3e}) with $G \leftrightarrow B$  in r.h.s.\
of~(\ref{3a}) we get for $B = \sum_{k=1}^{n} B_{k}$
\begin{equation}
M \{B, G \} = B + B \bullet M \left\{\dots
M\{M\{M\{G,B_{1}\}, B_{2}\}, B_{3} \},\dots,
B_{n}\right\}
\bullet B\,. \label{3b}
\end{equation}

The formulae~(\ref{3e}) and~(\ref{3b}) may provide useful schemes
of approximation for subdivision of Green functions and
irreducible Bethe-Salpeter kernels. Of course in principle these
two schemes can be combined. Let us mention that approaches with
suitable divisions of $G$ and/or $B$ have already been useful
both in the case of QED applications~\cite{Sapirstein} and for
deriving rigorous results concerning threshold behaviour in
QFT~\cite{Bros3}. The essential relation found in~\cite{Bros3}
exhibits convenience of using half-truncated amplitudes and
kernels $B_{H}$, $M_{H}$:
\begin{eqnarray}
M_{H} = M\{B, G \} \bullet G\,,\qquad B_{H} = B \bullet G\,.
\end{eqnarray}
For this case formula~(\ref{2j}) yields $M_{H} = M\{B_{H}, 1 \}$.
Moreover, from eq.~(\ref{3a})
\begin{equation}
M\{1, B_{H} \} = 1 + M\{ B_{H}, 1 \}\,.\label{3c}
\end{equation}
Using division of $B_{H}$ in $M\{1, B_{H} \}$ together with
eqs.~(\ref{3f}) and~(\ref{3c}) we recover in our notation
Bros-Iagolnitzer relation (compare text
before~\cite[eq.~(5)]{Bros3}):
\begin{equation}
1 + M \bullet G = (1 + M'' \bullet G) \bullet (1 - B' \bullet G
\bullet (1 + M'' \bullet G ))^{-1}\,,
\end{equation}
where $M = M\{B, G \}$, $M'' = M \{B - B', G \}$.

\section{Arbitrary trilinear coupling scheme}\label{s4}

We shall pass now to a problem where the use of the multi-channel
approach itself can be clearly exhibited and discussed, namely to
the example of a theory with general trilinear couplings of
scalar fields. Models of this kind may have phenomenological
applications when combined with effective, QCD-inspired
couplings~\cite{Pich,Bando,Cahill...}.

The most general trilinear coupling of $N$ hermitian scalar fields
$\varphi_i$ ($i = 1, \dots, N$) is of the form:
\begin{equation}
\mathcal{L}_\mrm{int}(x) = \frac{1}{3!} \sum_{i,j,k=1}^{N} c_{ijk}
{:}\varphi_{i}(x) \varphi_{j}(x) \varphi_{k}(x){:}\label{4f}
\end{equation}
The totally symmetric symbol $c_{ijk}$ denotes the coupling
constants.

Let us start with the off-shell amplitude:
\begin{equation}
i M^\mrm{r}(i_2, q_2; i_4, q_4| i_1, q_1; i_3, q_3)
= (2 \pi)^{4} \delta^4 (q_{1}+q_{3}-q_{2}-q_{4})
\mathcal{M}_{i_3 i_4}^{i_1 i_2}(P_{\mathrm{tot}}, q_{13}, q_{24})
\end{equation}
with $P_{\mathrm{tot}} = q_{1}+q_{3} = q_{2}+q_{4}$ and $q_{ij} =
\frac{q_{i} - q_{j}}{2}$.

In the reduced form~(\ref{2e}) of the Bethe-Salpeter equation in
the ladder approximation, the matrix $G^\mrm{r}$ can be chosen to
be
\begin{equation}\label{4l}
G^\mrm{r}(a, k_1; b, k_2| a', k_1'; b', k_2')
= (2 \pi)^8 \delta_{a a'} \delta_{b b'} \Delta_a(k_1) \Delta_b(k_2)
\delta^4(k_1 - k_1') \delta^4(k_2 - k_2')
\end{equation}
with free propagators $\Delta_a(k) = i / [k^2 - m_a^2 + i
\varepsilon]$, cf.\ (\ref{2i}). The Bethe-Salpeter equation for
$\mathcal{M}_{i_3 i_4}^{i_1 i_2}(P_{\mathrm{tot}}, q_{13}, q_{24})$ then
reads
\begin{eqnarray}
\mathcal{M}_{i_3 i_4}^{i_1 i_2}(P_{\mathrm{tot}}, q_{13}, q_{24})
&=& \mathcal{B}_{i_3 i_4}^{i_1 i_2}(q_{13}, q_{24})
+ \sum_{j,k} \int \md^{4}q \, \mathcal{B}_{i_3 k}^{i_1 j}(q_{13}, q)
\mathcal{G}_{k}^{j}(P_{\mathrm{tot}}, q)\times\nonumber\\
&&\hphantom{\mathcal{B}_{i_3 i_4}^{i_1 i_2}(q_{13}, q_{24}) +
\sum_{j,k} \int}\!\times\mathcal{M}_{k i_{4}}^{j i_{2}}(P_{\mathrm{tot}}, q,
q_{24})\,,
\end{eqnarray}
where $\mathcal{B}_{i_3 k}^{i_1 j}(q_{13}, q) = \sum_i c_{i_1 i j}
c_{i_3 i k} / [(q_{13} - q)^2 - m_i^2 + i \varepsilon]$ is the
Born term, and $\mathcal{G}_{k}^{j}(P_{\mathrm{tot}}, q)$ corresponds to
diagonal part of~(\ref{4l}):
\begin{eqnarray}
\mathcal{G}_k^j(P_{\mathrm{tot}}, q)
&=& - \frac{i}{(2 \pi)^4} \Delta_j\left(\frac{P_{\mathrm{tot}}}{2} + q\right)
\Delta_k\left(\frac{P_{\mathrm{tot}}}{2} - q\right)\nonumber \\
& = & \frac{i}{(2 \pi)^{4}} \frac{1}{(P_{\mathrm{tot}} / 2 + q)^2 - m_j^2
+ i \varepsilon} \frac{1}{(P_{\mathrm{tot}} / 2 - q)^2 - m_k^2 + i
\varepsilon}\,.
\end{eqnarray}
In the c.m.\ frame ($P_{\mathrm{tot}} = (W, 0)$) the partial-wave
decomposition of $\mathcal{M}_{i_3 i_4}^{i_1 i_2}$ is given~by
\begin{eqnarray}
\mathcal{M}_{i_3 i_4}^{i_1 i_2}(P_{\mathrm{tot}}^{\mathrm{c.m.}}, q_{1 3},
q_{2 4}) &=& \frac{1}{4 \pi |\vec{q}_{1 3}| |\vec{q}_{2 4}|}
\sum_{\ell = 0}^\infty (2 \ell + 1) P_\ell(\cos \vartheta)
\times
\nonumber\\&&
	\hphantom{\frac{1}{4 \pi |\vec{q}_{1 3}| |\vec{q}_{2 4}|}
		\sum_{\ell = 0}^\infty}\!
\times\mathcal{M}_\ell^{i_1, i_3; i_2, i_4}(W, q_{1 3}^0, |\vec{q}_{1 3}|,
q_{2 4}^0, |\vec{q}_{2 4}|)\,,\qquad
\end{eqnarray}
where $\vartheta$ is the angle between $\vec{q}_{1 3}$ and
$\vec{q}_{2 4}$.

In what follows we shall simplify the notation, introducing an
index $I(r, s)$ characterizing the channel $(r, s)$, i.e.\
\begin{equation}
(r, s) \longrightarrow I(r, s) = 1, \dots, N^2\,,
\end{equation}
and define matrices $\msf{M}_{\ell I J}$, $\msf{B}_{\ell I J}$ and
$\msf{G}_{I J}$ by
\begin{eqnarray}
\msf{M}_{\ell I(i, j) I(k, l)}(W, q_{1 3}^0, |\vec{q}_{1 3}|,
q_{2 4}^0, |\vec{q}_{2 4}|) &=& \mathcal{M}_\ell^{i, j; k, l}
(W, q_{1 3}^0, |\vec{q}_{1 3}|, q_{2 4}^0, |\vec{q}_{2 4}|) \,, \\
\msf{B}_{\ell I(i, j) I(k, l)}(q_{1 3}^0, |\vec{q}_{1 3}|, q_{2
4}^0, |\vec{q}_{2 4}|) &=& - 2 \pi \sum_{r = 1}^N c_{i r k} c_{j
r l} Q_\ell(\beta_r(q_{1 3}^0, |\vec{q}_{1 3}|, q_{2 4}^0,
|\vec{q}_{2 4}|))\,, \qquad\label{4g}  \\
\msf{G}_{I(i, j) I(k, l)}(W, k^0, |\vec{k}|, q^0, |\vec{q}|) & =
& \delta_{I(i, j) I(k, l)} \delta(k^0 - q^0) \delta(|\vec{k}| -
|\vec{q}|) \mathcal{G}_j^i(P_{\mathrm{tot}}, q) \,.
\end{eqnarray}
In~(\ref{4g}) $Q_\ell$ is an associated Legendre
function of the second kind, and its variable
$\beta_r$~is
\begin{equation}
\beta_r(q_{1 3}^0, |\vec{q}_{1 3}|, q_{2 4}^0, |\vec{q}_{2 4}|) =
\frac{- (q_{1 3}^0 - q_{2 4}^0)^2 + \vec{q}_{1 3}^2 + \vec{q}_{2
4}^2 + m_r^2 - i \varepsilon}{2 |\vec{q}_{1 3}| |\vec{q}_{2 4}|}\,.
\end{equation}
If the $\circ$ product for two such matrices is defined explicitly
by
\begin{eqnarray}
(\msf{X} \circ \msf{Y})_{IJ}(k^0, |\vec{k}|, q^0, |\vec{q}|) &:=&
\sum_{K = 1}^{N^2} \int \md q^{\prime 0} \, \md |\vec{q}\,'| \,
\msf{X}_{IK}(k^0, |\vec{k}|, q^{\prime 0}, |\vec{q}\,'|)\times
\nonumber\\
&&\hphantom{\sum_{K = 1}^{N^2}
\int}\!\times\msf{Y}_{KJ}(q^{\prime 0}, |\vec{q}\,'|, q^0,
|\vec{q}|)\,,
\end{eqnarray}
then the Bethe-Salpeter equation for the partial wave amplitudes
is again of the form~(\ref{2e}),
\begin{equation}
\msf{M}_\ell = \msf{B}_\ell + \msf{B}_\ell \circ \msf{G}
\circ \msf{M}_\ell \,,
\end{equation}
with the solution
\begin{equation}
\msf{M}_\ell = \msf{B}_\ell \circ \frac{1}{1 - \msf{G} \circ
\msf{B}_\ell} = \msf{M}\{\msf{B}_\ell, \msf{G}\}\,.
\end{equation}
Here, one may split $\msf{B}_\ell$ according to~(\ref{4g}) into
terms corresponding to the exchange of particles of type $r$,
\begin{equation}
\msf{B}_\ell = \sum_{r = 1}^N \msf{B}_\ell^{(r)}\,,
\end{equation}
$\msf{B}_{\ell I(i, j) I(k, l)}^{(r)} = - 2 \pi c_{i r k} c_{j r
l} Q_\ell(\beta_r)$. In this way we are naturally led to a
decomposition of the form~(\ref{3b}). Thus, $\msf{M}_\ell$ can be
written as
\begin{equation}
\msf{M}_\ell = \msf{B}_\ell + \msf{B}_\ell \circ \msf{M}
\left\{\dots
\msf{M}\{\msf{M}\{ \msf{G}, \msf{B}_\ell^{(1)}\},
\msf{B}_\ell^{(2)}\}, \dots, \msf{B}_\ell^{(N)}\right\} \circ
\msf{B}_\ell\,.
\end{equation}
This form of the solution may be useful for numerical studies of
$\msf{M}_\ell$ because the contributions of the various exchanged
particles are now disentangled in a systematic~way.

We will in the following focus our interest on Regge
trajectories near $\ell = - 1$; such a choice seems
sound once we recall that studies of this singularity in
$\varphi^3$ theory (i.e.\ only $c_{111} \neq 0$ in
\eq{4f}) often were and still are a point of departure
for studies of the diffractive region in QCD (see,
e.g.~\cite{Bartels}). In our discussion we shall not go
beyond the approximation of leading logarithms which
means that --- in analogy with the one-channel
case~\cite{Lee} --- $Q_\ell$ can be replaced by its
leading $\ell$-plane singularity, $Q_\ell(\beta)
\rightarrow 1/( \ell + 1)$. In this case, the matrix
$\msf{B}_\ell$ becomes constant,
\begin{eqnarray}
\msf{B}_{\ell I(i, j) I(k, l)}(k^0, |\vec{k}|, q^0, |\vec{q}|)
\longrightarrow - \frac{2 \pi}{\ell + 1} \msf{C}_{ I(i, j) I(k, l)}
&=& - \frac{2 \pi}{\ell + 1} \sum_{r = 1}^N c_{i r k} c_{j r l}
\nonumber\\ &=& \mbox{const}\,,
\end{eqnarray}
so that
\begin{equation}\label{4h}
\msf{M}_\ell = - 2 \pi \msf{C} \circ \frac{1}{\ell + 1 + 2 \pi
\msf{G} \circ \msf{C}}\,.
\end{equation}
Since $\msf{C}$ does not depend on the variables that are
integrated by the $\circ$ product,~(\ref{4h}) simplifies to give
a usual matrix equation: one finds
\begin{equation}
\msf{C} \circ \msf{G} \circ \msf{C} = - \frac{1}{2 \pi} \msf{C}
\msf{F} \msf{C}
\end{equation}
and so on, with the diagonal matrix $\msf{F}$ given by
\begin{equation}
\msf{F}_{I(i, j) I(k, l)}(W) = - 2 \pi  \delta_{I(i, j) I(k, l)}
\int \md q^0 \, \md |\vec{q}| \, \mathcal{G}_j^i(P_{\mathrm{tot}}, q)
\equiv \delta_{I(i, j) I(k, l)} F_{I(i, j)}(W)\,.
\end{equation}
Therefore, the solution of the Bethe-Salpeter equation is
\begin{equation}\label{4i}
\msf{M}_\ell = - 2 \pi \msf{C} \frac{1}{\ell + 1 - \msf{F}
\msf{C}}\,.
\end{equation}
Here everything is to be understood as a matrix relation. We will
in the following study the high-energy behaviour of the amplitude
in the crossed channel. To do this, we define the symmetric matrix
$\hat{\msf{C}} = \sqrt{\msf{F}} \msf{C} \sqrt{\msf{F}}$ and
write~(\ref{4i}) in the form
\begin{equation}
\msf{M}_\ell = - 2 \pi \sqrt{\msf{F}}^{- 1} \hat{\msf{C}}
\frac{1}{\ell + 1 - \hat{\msf{C}}} \sqrt{\msf{F}}^{- 1}\,.
\end{equation}
We shall limit our considerations to $P_{\mathrm{tot}}^{2} \le 0$
so that $F_{IJ}$ comes out to be real and positive. Since
$\hat{\msf{C}}$ is symmetric, we can diagonalize it: let
$\msf{v}^{[i]}$ denote the normalized eigenvectors,
\begin{equation}
\hat{\msf{C}} \msf{v}^{[i]} = \lambda^{[i]} \msf{v}^{[i]}\,,\qquad
i = 1, \dots, N^2\,,
\end{equation}
then
\begin{equation}
\msf{M}_\ell = - 2 \pi {\sqrt{\msf{F}}}^{- 1} \left( \sum_i
\frac{\msf{v}^{[i]} \lambda^{[i]} \msf{v}^{[i] \dagger}}{\ell + 1
- \lambda^{[i]}} \right) {\sqrt{\msf{F}}}^{- 1}\,. \label{4j}
\end{equation}
Thus, the Regge trajectories are given by
\begin{equation}
\alpha^{[i]} = \ell_{\mathrm{pole}} = \lambda^{[i]} - 1\,.
\end{equation}
With~(\ref{4j}) we are ready to use the
Mandelstam-Sommerfeld-Watson
transformation~\cite{Lee,Collins,Mandelstam} to obtain the
high-energy amplitude in the crossed channel ($t = (q_1 + q_3)^2
= W^2$ fixed, $s = (q_1 - q_2)^2 \rightarrow \infty$), with the
result
\begin{eqnarray}
\msf{M}_{\mathrm{poles}} &\stackrel{{s \longrightarrow \infty,}
\atop {t \, \mathrm{fixed}}}{=}& \frac{1}{s} {\sqrt{\msf{F}}}^{- 1}
\left( \sum_i \msf{v}^{[i]} \lambda^{[i]} s^{\lambda^{[i]}}
\msf{v}^{[i] \dagger} \right) {\sqrt{\msf{F}}}^{- 1}\nonumber\\ &=&
\frac{1}{s} {\sqrt{\msf{F}}}^{- 1} \left( \hat{\msf{C}} \exp
(\hat{\msf{C}} \ln s) \right) {\sqrt{\msf{F}}}^{- 1}\,.
\end{eqnarray}

The states $\msf{v}^{[i]}$ are eigenstates of the hermitian matrix
$\hat{\msf{C}}$, thus they are orthogonal. This is, however, no
longer true for the states $\sqrt{\msf{F}}^{- 1} \msf{v}^{[i]}$.
As the dual base for these states we introduce the states
\begin{equation}
\msf{u}^{[i]} = \sqrt{\msf{F}} \msf{v}^{[i]} \qquad \mbox{with}
\qquad \msf{u}^{[i] \dagger} \msf{F}^{- 1} \msf{u}^{[j]} =
\delta_{i j}\,.
\end{equation}
The only matrix elements of $\msf{M} = \msf{M}_{\mathrm{poles}}$
in this base that are non zero are $ \msf{u}^{[i] \dagger}
\msf{M} \msf{u}^{[i]} , i = 1, \dots, N^2 $. The high-energy
behaviour of each of these matrix elements is governed by only
one trajectory,
\begin{equation}\label{4a}
\msf{u}^{[i] \dagger} \msf{M} \msf{u}^{[j]} = \delta_{i j}
\lambda^{[i]} s^{\lambda^{[i]} - 1}\,,
\end{equation}
although the states $\msf{u}^{[i]}$ are in general not orthogonal.
Equation~(\ref{4a}) leads to the following sum rules for the
scattering amplitudes:
\begin{equation}\label{4b}
\sum_{I, J = 1}^{N^2} \msf{v}_I^{[i]} \msf{v}_J^{[j]} \sqrt{F_I F_J}
\msf{M}_{I J} = \delta_{i j} \lambda^{[i]} s^{\lambda^{[i]} - 1}\,.
\end{equation}
Note that some of these relations simplify whenever
$\lambda^{[i]} = 0$. In this case,
\begin{equation}
\msf{M} \msf{u}^{[i]} = 0\,, \qquad \mbox{for}\qquad
\lambda^{[i]} = 0
\end{equation}
(or equivalently $\msf{C} \msf{u}^{[i]} = 0$), i.e.\ states
corresponding to fixed Regge singularities are transparent in our
approximation.

As an illustration, let us consider the interaction
\begin{equation}
\mathcal{L}_{\mathrm{I}} =  \frac{g_1}{2!} \, {:} \varphi^2
\sigma {:} + \frac{g_2}{3!} \, {:} \sigma^3 {:}\,, \label{4k}
\end{equation}
which is a special case of~(\ref{4f}) for $N = 2$, $\varphi_1 =
\varphi$, $\varphi_2 = \sigma$, and $c_{112} = c_{121} = c_{211}
= g_1$, $c_{222} = g_2$, $c_{111} = c_{122} = c_{212} = c_{221} =
0$. Let $I(1, 1) = 1$, $I(2, 2) = 2$, $I(1, 2) = 3$, and $I(2, 1)
= 4$, then the matrix $\msf{C}$ is block diagonal due to the
symmetry of the total lagrangian with respect to $\varphi
\rightarrow - \varphi$,
\begin{eqnarray}
\msf{C} = \pmatrix{\msf{C}' & 0 \cr 0 & \msf{C}''},\qquad
\msf{C}' = \pmatrix{g_1^2 & g_1^2 \cr g_1^2 & g_2^2},\qquad
\msf{C}'' = \pmatrix{g_1 g_2 & g_1^2 \cr g_1^2 & g_1 g_2},
\end{eqnarray}
and $\msf{F}$ is diagonal with $F_{4} = F_{3}$, so the matrix
$\hat{\msf{C}}$ is
\begin{eqnarray}
\hat{\msf{C}} = \pmatrix{\hat{\msf{C}}' & 0 \cr 0 &
\hat{\msf{C}}''},\qquad
\hat{\msf{C}}' = \pmatrix{F_1 g_1^2 & \sqrt{F_1 F_2} g_1^2 \cr
\sqrt{F_1 F_2} g_1^2 & F_2 g_2^2},\qquad
\hat{\msf{C}}'' = F_3 \msf{C}''\,.
\end{eqnarray}
Diagonalizing $\msf{\hat{C}}'$, $\msf{\hat{C}}''$ we get
eigenvalues
\begin{equation}
\lambda^{[1, 2]} = \frac{1}{2} \left( F_1 g_1^2 + F_2 g_2^2 \pm
\sqrt{(F_1 g_1^2 - F_2 g_2^2)^2 + 4 F_1 F_2 g_1^4} \right)
\end{equation}
of $\msf{\hat{C}}'$ and
\begin{equation}
\lambda^{[3,4]} = F_3 g_1 (g_2 \pm g_1)
\end{equation}
of $\msf{\hat{C}}''$. The corresponding eigenvectors are
\begin{eqnarray*}
\msf{v}^{[1]} &=& \frac{1}{\sqrt{d}} \pmatrix{\sqrt{a} \cr
\sqrt{b} \cr 0 \cr 0},\qquad \msf{v}^{[2]} = \frac{1}{\sqrt{d}}
\pmatrix{\sqrt{b} \cr - \sqrt{a}
\cr 0 \cr 0},\\
\msf{v}^{[3]} &=& \frac{1}{\sqrt{2}} \pmatrix{0 \cr 0 \cr 1 \cr
1},\qquad \msf{v}^{[4]} = \frac{1}{\sqrt{2}} \pmatrix{0
\cr 0
\cr 1 \cr - 1},
\end{eqnarray*}
with $d = \sqrt{(F_1 g_1^2 - F_2 g_2^2)^2 + 4 F_1 F_2 g_1^4}$, $a
= (d + F_1 g_1^2 - F_2 g_2^2) / 2$, and $b = (d - F_1 g_1^2 + F_2
g_2^2) / 2$. Out of the ten relations described by~(\ref{4b}),
four are trivial due to the fact that $\msf{M}$ is block
diagonal. The remaining relations read (using the symmetry of
$\msf{M}$)
\begin{eqnarray}
a F_1 \msf{M}_{\varphi, \varphi \rightarrow \varphi, \varphi}
+ 2 F_1 F_2 g_1^2 \msf{M}_{\varphi, \sigma \rightarrow
\varphi, \sigma} + b F_2 \msf{M}_{\sigma, \sigma \rightarrow
\sigma, \sigma} &=& (\lambda^{[1]} - \lambda^{[2]})
\lambda^{[1]}s^{\lambda^{[1]} - 1}\,,
\nonumber\\
b F_1 \msf{M}_{\varphi, \varphi \rightarrow \varphi, \varphi}
- 2 F_1 F_2 g_1^2 \msf{M}_{\varphi, \sigma \rightarrow
\varphi, \sigma} + a F_2 \msf{M}_{\sigma, \sigma \rightarrow
\sigma, \sigma} &=& (\lambda^{[1]} - \lambda^{[2]}) \lambda^{[2]}
s^{\lambda^{[2]} - 1}\,,\nonumber\\
\msf{M}_{\varphi, \varphi \rightarrow \sigma, \sigma} + 2
\msf{M}_{\varphi, \sigma \rightarrow \sigma, \varphi} +
\msf{M}_{\sigma, \sigma \rightarrow \varphi, \varphi} &=& 2 g_1
(g_1 + g_2) s^{\lambda^{[3]} - 1}\,,\nonumber\\
\msf{M}_{\varphi, \varphi \rightarrow \sigma, \sigma}
- 2 \msf{M}_{\varphi, \sigma \rightarrow \sigma, \varphi}
+ \msf{M}_{\sigma, \sigma \rightarrow \varphi, \varphi}
&=& 2 g_1 (g_2 - g_1) s^{\lambda^{[4]} - 1}\,,\nonumber\\
F_1 g_1^2 \msf{M}_{\varphi, \varphi \rightarrow \varphi,
\varphi} + (F_2 g_2^2 - F_1 g_1^2) \msf{M}_{\varphi, \sigma
\rightarrow \varphi, \sigma} - F_2 g_1^2 \msf{M}_{\sigma,
\sigma \rightarrow \sigma, \sigma} &=& 0\,,\nonumber\\
\msf{M}_{\varphi, \varphi \rightarrow \sigma, \sigma} -
\msf{M}_{\sigma, \sigma \rightarrow \varphi, \varphi} &=&
0\,,\label{4d}
\end{eqnarray}
where, e.g.\ $\msf{M}_{\varphi, \sigma \rightarrow \varphi,
\sigma}$ is short for $\mathcal{M}((\vec{p}_1, \varphi),
(\vec{p}_2, \sigma) \rightarrow (\vec{p}_3, \varphi), (\vec{p}_4,
\sigma))$.

Of course, these relations are only valid in the Regge limit $s
\rightarrow \infty$, $t$ fixed, and apply only to Regge
behaviour, so that a vanishing combination of amplitudes means
that this combination does not exhibit Regge behaviour.

In the example, eigenvalues $\lambda^{[i]}$ can vanish only for
$g_2 = \pm g_1$. If, say, $g_2 = g_1$ (i.e.\ $\lambda^{[1]} =
g_1^2 (F_1 + F_2)$, $\lambda^{[2]} = 0$, $\lambda^{[3]} = 2 g_1^2
F_3$, $\lambda^{[4]} = 0$), the relations~(\ref{4d}) simplify~to
\begin{eqnarray}
\msf{M}_{\varphi, \varphi \rightarrow \varphi, \varphi}
&=& \msf{M}_{\sigma, \sigma \rightarrow \sigma, \sigma}
= \msf{M}_{\varphi, \sigma \rightarrow \varphi, \sigma}
= g_1^2 s^{\lambda^{[1]} - 1}\,,\nonumber\\
\msf{M}_{\varphi, \varphi \rightarrow \sigma, \sigma}
&=& \msf{M}_{\sigma, \sigma \longrightarrow \varphi, \varphi}
= \msf{M}_{\varphi, \sigma \longrightarrow \sigma, \varphi}
 =  g_1^2 s^{\lambda^{[3]} - 1}\,.\label{4e}
\end{eqnarray}
If moreover, we assumed $m_{\sigma} = m_{\varphi}$, then
$\lambda^{[1]} = \lambda^{[3]}$ and all the amplitudes standing
on the l.h.s.\ of~(\ref{4e}) are the same. This symmetry can be
traced back to the fact that for $g_{1} = g_{2} \equiv g$ we can
rewrite eq.~(\ref{4k}) as a sum of two separate $\chi_{\pm}^{3}$
terms
\begin{equation}
\mathcal{L}_\mrm{int}(g_{1} = g_{2} \equiv g)
= \frac{g}{2} \frac{1}{3!} \left[ {:}\chi_{+}^{3}{:}
+ {:}\chi_{-}^{3}{:} \right],
\end{equation}
where $\chi_{\pm} = \sigma \pm \varphi$. Thus, in the special case
$m_{\sigma} = m_{\varphi}$ the total lagrangian can be separated
into two of identical form and hence the complete symmetry
follows.

\section{Concluding remarks}\label{s5}

The general results of sections~\ref{s2} and \ref{s3} provide a
clear and general scheme of dealing with various approximations
in the multi-channel Bethe-Salpeter equation. Moreover, the
results derived in section~\ref{s4} show that the multi-channel
Bethe-Salpeter formalism is tractable and leads to non-trivial
physical relations, e.g.\ sum rules for scattering amplitudes. In
our opinion, future applications may include the treatment of the
$e^+ e^-$ and $\gamma \gamma$ channels of parapositronium, and
the Bethe-Salpeter analysis of hadronic multi-channel processes
hitherto treated not fully relativistic. In the latter case
theoretical input could be provided by effective couplings
derived from QCD~\cite{Pich,Bando,Cahill...}.

\acknowledgments

We are indebted to Professors Jacques Bros, Wojciech
Kr\'olikowski and Hagop Sazdjian for the stimulating
correspondence and discussions.

\end{document}